\documentclass[aps,prd,preprint,preprintnumbers,groupedaddress,nofootinbib]{revtex4-1}
\usepackage{amsmath,amssymb,amsfonts,graphicx,color}

\newcommand{\be}{\begin{equation}}
\newcommand{\ee}{\end{equation}}
\newcommand{\bea}{\begin{eqnarray}}
\newcommand{\eea}{\end{eqnarray}}

\newcommand{\mc}{\mathcal}

\newcommand{\beqa}{\begin{eqnarray}}
\newcommand{\eeqa}{\end{eqnarray}}

\newcommand{\vo}{\mathcal{V}}

\newcommand{\nn}{\nonumber}

\newcommand\fverb{\setbox\fverbbox=\hbox\bgroup\verb}
\newcommand\fverbdo{\egroup\medskip\noindent
			\fbox{\unhbox\fverbbox}\ }
\newcommand\fverbit{\egroup\item[\fbox{\unhbox\fverbbox}]}
\newbox\fverbbox

\begin{document}
\preprint{MIFPA 13-25}
\title{Power Suppression at Large Scales in String Inflation}
\author{Michele Cicoli}
\affiliation{Dipartimento di Fisica ed Astronomia, Universit\`a di Bologna, \\ via Irnerio 46, 40126 Bologna, Italy}
\email[]{mcicoli@ictp.it}
\affiliation{INFN, Sezione di Bologna, 40126 Bologna, Italy}
\affiliation{Abdus Salam ICTP, Strada Costiera 11, Trieste 34014, Italy}

\author{Sean Downes}
\email[]{sddownes@physics.tamu.edu}
\author{Bhaskar Dutta}
\email[]{dutta@physics.tamu.edu}
\affiliation{Mitchell Institute for Fundamental Physics and Astronomy \\ Department of Physics and Astronomy, Texas A\&M University, \\ College Station, TX 77843-4242, USA}

\begin{abstract}
We study a possible origin of the anomalous suppression of the power spectrum at large angular scales in the cosmic microwave background
within the framework of explicit string inflationary models where inflation is driven by a closed string modulus parameterizing
the size of the extra dimensions. In this class of models the apparent power loss at large scales is caused by the
background dynamics which involves a sharp transition from a fast-roll power law phase to a
period of Starobinsky-like slow-roll inflation. An interesting feature of this class of string inflationary models
is that the number of e-foldings of inflation is inversely proportional to the string coupling
to a positive power. Therefore once the string coupling is tuned to small values in order to trust string perturbation theory,
enough e-foldings of inflation are automatically obtained without the need of extra tuning.
Moreover, in the less tuned cases the sharp transition responsible for the power loss takes place just before
the last 50-60 e-foldings of inflation. We illustrate these general claims in the case of Fibre Inflation
where we study the strength of this transition in terms of the attractor dynamics,
finding that it induces a pivot from a blue to a redshifted power spectrum which can explain the apparent large scale power loss.
We compute the effects of this pivot for example cases and demonstrate how magnitude and duration of this effect depend on model parameters.
\end{abstract}

\maketitle
\tableofcontents
\newpage

\section{Introduction}

The major ideas of the inflationary paradigm have received further confirmation by the recent results from the Planck Collaboration~\cite{Ade:2013zuv}. The new results have provided measurements of the cosmic microwave background (CMB) with increased sensitivity.
An interesting feature of the Planck measurements is the observation of an anomalously low power of
the thermal fluctuations at large angular scales. This anomaly was already present in the observations of the Cosmic Background Explorer~\cite{Bond:1998zw}
and was observed with a greater accuracy by Wilkinson Microwave Anisotropy Probe (WMAP)~\cite{Bennett:2003bz}.

Given that inflation predicts a roughly flat spectrum of primordial perturbations --- and the subsequent evolution of these perturbations does little to qualitatively modify the largest scales --- the fluctuations probed by the lowest multipole moments should be roughly scale-invariant.
The ``anomaly'' refers to the fact that the lowest moments of the angular power spectrum are observed to be lower than expected.

The origin of this anomaly is not concluded to be primordial yet; there exist questions regarding statistical analysis and secondary anisotropies.
If this anomaly is indeed primordial in nature, then the physics associated to this power suppression must have acted just prior to the observationally relevant number of e-foldings before the end of inflation. This remains an open problem, and it is worthwhile to understand its origin and general features in inflationary models. In this paper we investigate this anomaly in the context of inflationary models arising from string theory.

The inflation paradigm involves a scalar field, dubbed the ``inflaton'', which drives a
period of exponential expansion of the spacetime metric.
This field evolves slowly during the inflationary stage with the dynamics determined by a fairly flat potential.
This scenario is supported by all cosmological observations and is referred to as slow-roll inflation~\cite{Linde1982389,PhysRevLett.48.1220}.

It has already been observed that a deviation from the slow-roll behavior during a pre-inflationary phase
just prior to horizon exit can generate a loss of power at large scales by introducing
an energy cut-off in the power spectrum that modifies the density fluctuations~\cite{Burgess:2002ub,Contaldi:2003zv,Cline:2003ve,BasteroGil:2003bv,
Shankaranarayanan:2004dk,Buchel:2004df,Jain:2008dw,Schwarz:2009sj,Ramirez:2012gt,Dudas:2012vv,Lello:2013mfa}.
The presence of this cut-off could arise from an early ``stringy'' epoch,
an initial period of matter or radiation domination,
an oscillating scalar field coupled to the inflaton,
or large radiative corrections to the effective inflaton potential.\footnote{An enhancement/suppression of
the power spectrum at large scales might also be due to first order corrections to the scalar fluctuations
coming from quantum gravity effects \cite{Kamenshchik:2013msa}.}
These are all models of ``just enough'' inflation, where the amount of inflation is just the one necessary to solve the problems
of the standard cosmological scenario.

If the power suppression at large scales is indeed a physical effect, this raises a ``coincidence problem'':
why did inflation last just long enough to solve the flatness and horizon problems?
In principle, there is no reason why inflation should have lasted only for 50-60 e-foldings but not more.
On the other hand, if the slow-roll evolution gave rise to a number of e-foldings larger than
50-60, we should not see any suppression of the power spectrum at the longest wavelength modes.

Recently, it was shown that a deviation from slow-roll occurs naturally in inflection point scenarios
where the inflaton changes its trajectory from fast to slow-roll while approaching the inflection point~\cite{Downes:2012gu}.
This transition from chaotic to inflection point inflation causes the spectrum to pivot from blue to red,
and can be tuned to occur at the onset of fifty to sixty e-foldings of inflation.

However, a transition can occur also from a power law to a Starobinsky-like potential and generate a loss of power at large scales.
In this paper we shall argue that there exist promising inflationary models with this feature arising from string theory~\cite{Burgess:2013sla}.
These are models where the inflaton is a closed string modulus parameterizing the size of the extra dimensions,
namely a K\"ahler modulus which is the volume of a divisor of the internal Calabi-Yau manifold \cite{Cicoli:2011zz}.
Possible models within this scenario include: K\"ahler Moduli \cite{Conlon:2005jm}, Fibre \cite{Cicoli:2008gp} and Poly-Instanton Inflation \cite{Cicoli:2011ct}.
These models differ from each other due to the different, perturbative or non-perturbative, effects used to develop the inflationary potential,
and the different topological nature of the inflaton, but share the same solution to the $\eta$-problem based on an approximate shift
symmetry of the K\"ahler potential caused by the so-called ``extended no-scale structure'' \cite{Cicoli:2007xp}.
Moreover, this class of models leads to the rather universal predictions of a spectral index $n_s$ in the observed window,
a non-detectable tensor to scalar ratio $r$ (with some differences between the three models) and a lack of observable non-Gaussianity~\cite{Burgess:2013sla}.

This perfect agreement with the Planck observations is due to the fact that in each of these models
the inflationary potential has the same shape of the $R+R^2$-model \cite{Starobinsky:1980te} even if it has a different microscopic derivation.
In particular, in this class of models, the inflaton does not develop its potential by balancing different terms in the same expansion
but by competing terms in different expansions: $\alpha'$ and $g_s$ perturbative effects on top of non-perturbative corrections to the effective action.
In this sense, these models are more robust than the Starobinsky scenario.

However, for large values of the inflaton, stringy corrections spoil the flatness of the potential,
inducing a natural transition from an early period of fast-roll evolution to a later stage of standard slow-roll inflationary dynamics.
This transition between inflationary trajectories can differentiate these stringy models from
standard $R+R^2$ scenarios, and be responsible for the loss of power at the largest scales
if it occurred just before the last 50-60 observed e-foldings of inflation.

As we have already pointed out, this ``coincidence problem'' requires in general a specific tuning of initial conditions.
However, in these string inflationary models, the flatness of the potential ---
and therefore the number of redshifted e-foldings of inflation $N_e$ ---
is tied to the size of the string coupling $g_s$. We shall show that $N_e\propto g_s^{-k}$ with $k>0$,
implying that larger numbers of e-foldings correspond to more fine-tuned situations with a tiny value
of the string coupling which is the set by the underlying choice of background flux quanta.
Therefore, the requirement of trusting string perturbation theory for $g_s\ll 1$, automatically
guarantees enough e-foldings to solve the flatness and horizon problems,
preferring values of $N_e$ as close as possible to its minimal value compatible with
cosmological observations.

In this paper, we shall illustrate these general claims in the particular case of Fibre Inflation
and explicitly calculate the amount of power loss in terms of model parameters. We will see that a large drop of power as observed in the data is achievable for reasonable values of model parameters.

This paper is organized as follows: in Sec.~\ref{sec:two} we discuss slow-roll inflation and density perturbations. In Sec.~\ref{sec:three},
we summarize the Fibre Inflation model. In Sec.~\ref{ch:lowpower} we calculate the power spectrum at large scales in the Fibre inflation model. Finally, in Sec.~\ref{sec:five} we conclude with a discussion of our results.

\section{A Slow-Roll Primer}
\label{sec:two}

This section is a brief review of the aspects of slow-roll inflation required for the study of the large scale power loss from string-inspired models.
In addition to establishing conventions and notation, we emphasize dynamical features and background solutions which we will employ in our discussion.
To facilitate the discussion on power loss, we emphasize the possibility of a pivot from a blue to a red-shifted spectrum.
In our analysis we employ natural units where $c=\hbar = M_P =1$.

\subsection{Evolution of the background}

We study a single scalar field $\varphi$ minimally coupled to gravity under the cosmological assumptions of spatial homogeneity and isotropy. We employ the spatially-flat Friedmann-Lema\^itre-Robertson-Walker metric, whose line element is:
$$ds^2 = -dt^2 + a^2(t)\left(dx^2 + dy^2 + dz^2\right).$$
The inclusion of non-vanishing spatial curvature is straightforward and does not qualitatively affect the results below. The well-known field equations for the gravity-scalar system are:
\be
3H^2=\frac{1}{2}\dot{\varphi}^2 + V\,,\qquad\quad\dot{H}=-\frac{1}{2}\dot{\varphi}^2\,,
\qquad\quad
\ddot{\varphi} + 3H\dot{\varphi} + V_{\varphi} =0\,.
\label{fieldEQs}
\ee
Dots represent time derivatives, $H=\dot{a}/a$, and a subscript $\varphi$ denotes a partial derivative with respect to the $\varphi$ field.
It is convenient to parameterize time with the number of e-foldings of expansion $N_e=\ln a$.
In such a case, $\dot{\varphi} = \varphi^{\prime}H$, where primes denote derivatives with respect to $N_e$. In this variable the Hubble parameter is:
\be
H^2 = \frac{V}{3(1-\frac{1}{6}\varphi^{\prime 2})}\,,
\ee
and the third field equation in \eqref{fieldEQs} becomes:
\be
\label{eqn:coKGN}
\varphi^{\prime\prime}=\frac{1}{2}(\varphi^{\prime}+\sqrt{6})(\varphi^{\prime}-\sqrt{6})\left(\varphi^{\prime}+\frac{V_{\varphi}}{V}\right).
\ee
From this perspective, the would-be singular solution to \eqref{eqn:coKGN}:
\be
\label{eqn:newSR}
\varphi^{\prime}\approx -\frac{V_{\varphi}}{V}= - \partial_\varphi \ln V\,,
\ee
is a dynamical attractor which describes slow-roll trajectories.
In fact, using the standard definition of the slow-roll parameters $\epsilon$ and $\eta$:
\be
\label{eqn:oldSR}
\epsilon \equiv \frac{1}{2}\Big(\frac{V_{\varphi}}{V}\Big)^2\qquad\text{and}\qquad
\eta \equiv\frac{V_{\varphi\varphi}}{V}\,,
\ee
the relation (\ref{eqn:newSR}) implies:
\be
\varphi^{\prime\prime} \approx -\left(\eta-2\epsilon \right)\varphi^\prime\,,
\ee
which is a singular solution to \eqref{eqn:coKGN} only if the
usual slow-roll conditions are satisfied: $\epsilon \ll 1$ and $|\eta| \ll 1$.\footnote{For a broader discussion in this context see \cite{Downes:2012xb}.}
Phrased in terms of kinematic variables, the slow-roll approximation is valid if the quantity $\Xi$, defined as:
\be
\label{eqn:defineXi}
\Xi \equiv \frac{\varphi^{\prime\prime}}{\varphi^{\prime}}\,,
\ee
is very close to zero.
If $V_{\varphi}/V$ happens to be a constant, then \eqref{eqn:newSR} is an exact solution to the equations of motion. This can be realized with exponential potentials, for example. Such potentials can induce a period of power law inflation. This will play an important r\^ole in our discussion of power loss below.

Lets us make a final comment about accelerated expansion. The so-called deceleration parameter $q$ obeys:
\be
\label{eqn:decel}
(-q) = \frac{\ddot{a}a}{\dot{a}^2} = 1-\frac{\dot{H}}{H^2} = 1-\frac{1}{2}\phi^{\prime 2}\,.
\ee
Therefore, a universe dominated by a scalar condensate has accelerated expansion only if $|\phi^{\prime}|<\sqrt{2}$. Note that an artifact of parameterizing time with $N_e$ is a bound on $|\phi^{\prime}|\leq \sqrt{6}$, which is saturated during the case of kinetic energy domination.

\subsection{The curvature perturbations}

Linear perturbations around the ($a$,$\varphi$)-system define primordial quantum fields whose observed effects modulate the otherwise spatially uniform background. To linear order, the curvature perturbation $\mathcal{R}(t,\vec{r})$, can be thought of as a scalar field representing local deviations from the background value of $N_e$.
When computing the perturbations, we work in the comoving gauge which fixes metric and $\varphi$ perturbations to be:
\be
g_{ij} = a^2\left[\left(1-2\mathcal{R}\right)\delta_{ij}+h_{ij}\right]\qquad\text{and}\qquad\delta\varphi=0\,.
\ee
Here $h_{ij}$ is transverse-traceless and represents the primordial tensor modes.

To linear order, the perturbed field equations for the Fourier modes, $\mathcal{R}_k$, of $\mathcal{R}$ are:
\be
\label{eqn:Reqn}
\mathcal{R}^{\star\star}_{\vec{k}} + 2aH\mathcal{R}^{\star}_{\vec{k}} +\left[k^2 - 2H(1+\Xi)\right]\mathcal{R}_{\vec{k}} = 0\,,
\ee
where $k=|\vec{k}|$ and stars denote derivatives with respect to conformal time $\eta(t) = \int_{-\infty}^{t}\frac{d\tau}{a(\tau)}$.
The Mukhanov-Sasaki variable $\mathcal{Q}=a\mathcal{R}$ brings \eqref{eqn:Reqn} into a more tractable (Liouville) form:
\be
\label{eqn:Qeqn}
\mathcal{Q}_k^{\star\star} + \left[k^2 -(aH)^2\left((2+\Xi-\frac{1}{2}\phi^{\prime 2})(1+\Xi)+\Xi^{\prime}\right)\right]\mathcal{Q}_k=0\,.
\ee
The solutions to \eqref{eqn:Qeqn} can be written in terms of spherical Hankel functions. For a consistent, Minkowski-like vacuum we consider only \textit{outgoing} modes, which specifies one of the boundary values. The normalized wavefunctions are then:
\be
\label{eqn:BigQ}
\mathcal{Q}_k = \frac{k\eta\, h_{\nu}^{(1)}(k\eta)}{\sqrt{2k}}\,,
\ee
where:
\be
\label{eqn:nudef}
\nu(1+\nu) = (aH\eta)^2\left[(2+\Xi-\frac{1}{2}\phi^{\prime 2})(1+\Xi)+\Xi^{\prime}\right].
\ee
During slow-roll inflation, the prefactor $(aH\eta)^2$ is exponentially close to unity.
In terms of the spectral-tilt, $\nu$ will have a crucial r\^ole to play in Section~\ref{ch:lowpower}.

The power spectrum of primordial perturbations $\mathcal{P}(k)$ is related to the variance of $\mathcal{R}_k$:
\be
\label{eqn:PS}
\langle\mathcal{R}_{\vec{k}}\mathcal{R}_{\vec{k}^{\prime}}\rangle=(2\pi)^3\mathcal{P}(k)\delta^{(3)}(\vec{k}+\vec{k}^{\prime})\,.
\ee
The (scalar) spectral index is then defined to be:
\be
\label{eqn:ns}
n_s = 3-\frac{d\log\mathcal{P}(k)}{d\log k}\,,
\ee
so that $n_s=0$ corresponds to fluctuations around a quasi-de Sitter background:
\be
\mathcal{P}_{dS}(k) = \frac{H^4}{\dot{\phi}^2 k^3}\,,
\ee
i.e. one with an inverse cubic power for $k$.

Having reviewed the basics of inflation, we now describe the string-inspired context in which we perform our analysis of the pivoting spectral tilt.

\section{Fibre Inflation}\label{sec:three}

In this section we shall review the main features of Fibre Inflation \cite{Cicoli:2008gp}.

\subsection{The leading potential}
\label{3modK3noLoopCalc}

The simplest version of Fibre Inflation involves a Calabi-Yau three-fold with $h^{1,1}=3$ K\"ahler moduli, $\tau_i$, $i=1,2,3$,
and a volume of the form \cite{Cicoli:2011it}:
\be
 \vo = \alpha \left( \sqrt{\tau_1}\tau_2 - \gamma \tau_3^{3/2}\right)\,,  \label{hhh}
\ee
where the constants $\alpha $ and $\gamma $ are given in terms of
the triple intersections numbers as $\alpha = 1/\sqrt{2 \kappa_{122}}>0$ and $\gamma = \frac 23 \sqrt{\frac{\kappa_{122}}{k_{333}}}>0$.
We shall also work in the regime:
\be
\label{V0hier}
 \vo\simeq \alpha\sqrt{\tau_1}
 \; \tau_2 \gg \alpha\gamma\tau_3^{3/2} \gg 1 \,.
\ee

Let us consider the scalar potential computed using the
leading $\alpha'$ corrections to the K\"ahler potential, as well
as including non-perturbative corrections to the superpotential:
\be
\label{3}
 K = K_0 + \delta K_{(\alpha')} = -2 \ln \left(
 \vo + \frac{\hat{\xi}}{2} \right) \qquad \hbox{and}
 \qquad
 W = W_0 + A_3 e^{-a_3 T_3} \,,
\ee
where $W_0$ is the flux-generated tree-level superpotential, $T_3=\tau_3+{\rm i}b_3$, $A_3$ is an $\mc{O}(1)$ constant and:
\be
\hat{\xi}\equiv \frac{\xi}{g_s^{3/2}}=\frac{\zeta(3)\left(h^{1,2}-3\right)}{(2\pi)^3 g_s^{3/2}}
\qquad\text{and}\qquad a_3=\frac{2\pi}{N_3}\,,
\ee
with $N_3 \in \mathbb{N}$ and $h^{1,2}$ giving the number of complex structure moduli
stabilized by background fluxes at tree-level.
After minimizing with respect to the $T_3$-axion $b_3$ ($\langle b_3\rangle = \pi/a_3$),
the resulting F-term supergravity scalar potential looks like:
\be
 V_{\rm lead} = \frac{8 \, a_3^2 A_3^2}{3\alpha\gamma}
 \left( \frac{\sqrt{\tau_3}}{\vo} \right) e^{-2 a_3\tau_3}
 -4 |W_0| a_3 A_3 \left( \frac{\tau_3}{\vo^2}\right) \, e^{-a_3\tau_3}
 +\frac{3 \, \hat\xi |W_0|^2}{4 \vo^3} +V_{\rm up}\,,
 \label{ygfdo}
\ee
where we added also an uplifting contribution $V_{\rm up}$ to obtain a Minkowski vacuum.
Notice that, at this order of approximation, $V$ depends only on two of the three moduli: $V=V(\vo,\,\tau_3)$.
Therefore the combination of $\tau_1$ and $\tau_2$ orthogonal to $\vo$ is a flat direction,
resulting in a perfect candidate for an inflaton field.
On the other hand, the minimum for $\vo$ and $\tau_3$ is located at:
\be
 \langle \tau_3\rangle = \left(
 \frac{\hat\xi}{2\, \alpha \gamma } \right) ^{2/3}
 \qquad \hbox{and} \qquad
 \langle  \vo\rangle = \left( \frac{ 3 \,\alpha \gamma }{4 a_3 A_3}
 \right) |W_0| \, \sqrt{\langle\tau_3\rangle}
 \; e^{a_3 \langle \tau_3 \rangle }\,,  \label{x}
\ee
where the generic situation without any tuning of the underlying flux quanta gives $|W_0|\sim \mc{O}(1)$.
The remaining flat direction can be lifted at subleading order in a large volume expansion
by including string loop corrections to the K\"ahler potential.
These effects will develop a potential for the inflaton field.

\subsection{The subleading potential}
\label{bene}

The $\tau_1$ and $\tau_2$-dependent string loop corrections to the scalar potential can be estimated
to scale as \cite{Cicoli:2008gp}:
\be
 V_{\rm sub}\simeq \left(\frac{A}{\tau_1^2} -\frac{B}{\vo\sqrt{\tau_1}} +\frac{C\tau_1}{\vo^2}\right)\frac{|W_0|^2}{\vo^2},
 \label{eqn:prepots}
\ee
where the constants $A$, $B$ and $C$ are given in terms of the underlying parameters $g_s\ll 1$ and
$C_1^{KK}\sim C_2^{KK}\sim C_{12}^{W}\sim \mc{O}(1)$ as (see \cite{Cicoli:2008gp} for more details):
\be
 A=\left( g_s C_1^{KK}\right)^2>0\qquad
 B=4 \alpha C_{12}^{W}\qquad
 C= 2\,\left( \alpha g_s\,C_2^{KK}\right)^2>0 \,.
 \label{defC}
\ee
As we shall see later on, the term in (\ref{eqn:prepots}) proportional to $B$ is responsible
to drive the inflationary dynamics, while the term proportional to $A$ controls the end of inflation.
On the other hand, the term proportional to $C$ determines the power suppression at large angular scales.
Given that the expression (\ref{eqn:prepots}) is not the result of an exact computation of string scattering amplitudes,
but it is more an educated guess based on toroidal orientifold computations \cite{Berg:2007wt}
and the low-energy Coleman-Weinberg potential \cite{Cicoli:2007xp}, we shall consider a more
general form of the term proportional to $C$ in order to perform a more complete analysis of
the power loss at large scales. On dimensional grounds, we generalize:
\be
\frac{C\tau_1}{\vo^2}\qquad\longrightarrow\qquad\frac{C\tau_1}{\vo^2}\left(\frac{\tau_1}{\tau_2}\right)^{\frac{2n}{3}}
=\frac{C\tau_1^{1+n}}{\vo^{2+\frac{2n}{3}}}\,,
\label{generalize}
\ee
where we take $n>0$ to be a phenomenological parameter whose vanishing leads to the limit of \eqref{eqn:prepots}.
Physically, $n$ controls the leading behavior at large $\tau_1$.
As discussed below, the pivot from a blue to a red-shifted spectrum is very sensitive to $n$.

The minimisation condition for the generalized potential can be written as:
\be
x^2 - x - \left(1+n\right)\frac{\delta_0}{2}\left(\frac{\zeta}{x}\right)^{\frac{2n}{3}}=0\qquad\text{with}
\qquad x\equiv\frac{\zeta \vo}{\tau_1^{3/2}}\,,
\label{minimisation}
\ee
where the small parameters $\delta_0$ and $\zeta$ are defined as:
\be
\delta_0\equiv \frac{16 AC}{B^2}= 2 g_s^4 \left(\frac{C_1^{KK} C_2^{KK}}{C_{12}^W} \right)^2 \ll 1
\qquad\text{and}\qquad \zeta\equiv\frac{4\,A}{B}=g_s^2\,\frac{\left(C_1^{KK}\right)^2}{\alpha C_{12}^{W}}\ll 1\,.
\label{delta}
\ee
Given that $\delta_0 \simeq \mc{O}(g_s^4)$ and $\zeta\simeq \mc{O}(g_s^2)$,
both of these parameters are naturally much smaller than unity for $g_s\ll 1$,
i.e. in the regime where one can trust string perturbation theory.

In the $n=0$ case, (\ref{minimisation}) becomes:
\be
x^2-x-\frac{\delta_0}{2}=0\,,
\ee
which admits two solutions for $\delta_0\ll 1$: $x\simeq 0+\mc{O}(\delta_0)$ and $x\simeq 1+\mc{O}(\delta_0)$.
Focusing on positive values of $B$, i.e. positive values of $\zeta$, which in turn implies
positive values of $x$, the first solution has actually to be discarded since it becomes
negative once subleading corrections are taken into account: $x \simeq -\delta_0/2$.
We shall therefore focus on the second solution. Including also subdominant effects, we obtain:
\be
x \simeq 1+\frac{\delta_0}{2}\qquad\Leftrightarrow\qquad
 \langle \tau_1 \rangle
 \simeq \left(\zeta\vo \right)^{2/3}\left(1-\frac{\delta_0}{3}\right)\,.
 \label{tau1soln2}
\ee
In the general case with $n>0$, the third term in (\ref{minimisation}) keeps being a subleading effect
due to the smallness of $\delta_0$ and $\zeta$ and the fact that the solution for $x$ is around $x\simeq 1$.
By defining the new small parameter $\delta$ as:
\be
\delta \equiv \left(1+n\right)\delta_0 \zeta^{\frac{2n}{3}}\simeq
\mc{O}\left(g_s^{4\left(1+\frac{n}{3}\right)}\right)\ll 1  \quad\underset{n\to 0}{\longrightarrow} \quad\delta_0\,,
\label{gs}
\ee
the minimisation condition (\ref{minimisation}) can be rewritten as:
\be
x^2-x-\frac{\delta}{2}\,x^{-\frac{2n}{3}}=0\,.
\label{minGEN}
\ee
Parameterizing the solution in terms of an unknown constant $\lambda$ as $x = 1+ \lambda \delta$,
we can solve (\ref{minGEN}) exactly at $\mc{O}(\delta)$:
\be
x^2-x-\frac{\delta}{2}\,x^{-\frac{2n}{3}}= \left(\lambda-\frac{1}{2}\right)\delta+\mc{O}(\delta^2)
\qquad\Rightarrow\qquad \lambda=\frac{1}{2}\,.
\ee
The final solution for the VEV of $\tau_1$ is therefore:
\be
\langle \tau_1 \rangle
 \simeq \left(\zeta\vo \right)^{2/3}\left(1-\frac{\delta}{3}\right)\,.
\ee

\subsection{The canonically normalized inflaton potential}

In order to work out the scalar potential for the canonically normalized inflaton field,
we need to consider the kinetic terms which take the form (trading $\tau_2$ for $\vo$):
\be
 - \frac{{\cal L}_{\rm kin}}{\sqrt{-g}}
 =\frac{3}{8 \tau_1^2}\left( \partial \tau_1\right)^2
 +\frac{1}{2 \vo^2}\left( \partial \vo\right)^2 - \frac{1}{2 \tau_1 \vo}
 \,\partial \tau_1 \partial \vo \,.
 \label{LagKin}
\ee
In the previous expression we neglected terms involving derivative with respect to $\tau_3$ since these
contributions are suppressed by $1/\vo\ll 1$. The expression (\ref{LagKin}) can be put into canonical
form by performing the following transformation:
\be
\vo = e^{\sqrt{\frac32} \chi},\qquad
 \tau_1 = e^{\sqrt{\frac23} \chi + \frac{2}{\sqrt{3}} \phi}\,. \label{vol}
\ee
As shown in \cite{Cicoli:2010ha}, $\chi$ and $\phi$ are also eigenstates of the mass matrix
(up to a tiny dependence of $\vo$ on $\phi$ introduced by string loop effects).
The moduli $\vo$ and $\tau_3$ are heavier than $\tau_1$, and so they will reach their minima first.
In order to study the inflationary dynamics, we can therefore consider $\vo$ and $\tau_3$ as fixed
at their minima and $\tau_1$ slowly rolling toward its minimum.
In this situation, $\tau_1$ is given in terms of the canonically normalized inflaton $\phi$ as:
\be
\tau_1 = \langle\tau_1\rangle \,e^{\frac{2}{\sqrt{3}}\hat{\phi}}\simeq \zeta^{2/3}\langle\vo\rangle^{2/3}\left(1-\frac{\delta}{3}\right)\,e^{\frac{2}{\sqrt{3}}\,\hat\phi}\,,
\label{tau1norm}
\ee
where we shifted $\phi$ with respect to its minimum as $\phi=\langle\phi\rangle+\hat\phi$.

Plugging the expression (\ref{tau1norm}) in (\ref{eqn:prepots}),
and taking into account the generalization (\ref{generalize}),
the inflationary potential becomes (working to linear order in $\delta$):
\be \label{infpot}
 V_{\rm inf} =\frac{A |W_0|^2}{\zeta^{4/3}\langle\vo\rangle^{10/3}}\,\left[\left(1+\frac{2\delta}{3}\right)
 e^{-\frac{4}{\sqrt{3}}\, \hat\phi}-4\left(1+\frac{\delta}{6}\right)\,e^{-\frac{1}{\sqrt{3}}\,\hat\phi}
 +\frac{\delta}{\left(1+n\right)}\,e^{\frac{2}{\sqrt{3}}\left(1+n\right) \hat\phi} + C_{\rm up}\right],
\ee
where the uplifting contribution $C_{\rm up}$ has to be adjusted such that
$V_{\rm inf}(\langle \hat\phi \rangle=0) = 0$. This requires $C_{\rm up} = 3-\frac{\delta}{\left(1+n\right)}$.
The prefactor of the potential (\ref{infpot}) can be rewritten in terms of the mass of the inflaton field
around the minimum since:
\be
m_\phi^2 = \frac{\partial^2 V_{\rm inf}}{\partial \hat\phi^2}\Bigg|_{\hat\phi=0}
=\frac{4 A |W_0|^2}{\zeta^{4/3}\langle\vo\rangle^{10/3}}\left[1+\frac{7\delta}{6}\left(1+ \frac{2n}{7}\right) \right].
\ee
This gives the final expression for the inflationary potential:
\be \label{InfPot}
 V_{\rm inf} \simeq\frac{m_\phi^2}{4}\,
 \left[\left(1+\frac{2\delta}{3}\right)
 e^{-\frac{4}{\sqrt{3}}\, \hat\phi}-4\left(1+\frac{\delta}{6}\right)\,e^{-\frac{1}{\sqrt{3}}\,\hat\phi}
 +\frac{\delta}{\left(1+n\right)}\,e^{\frac{2}{\sqrt{3}}\left(1+n\right) \hat\phi} + 3-\frac{\delta}{\left(1+n\right)}\right].
\ee
The potential from \eqref{InfPot} is illustrated for various values of $\delta$ and $n$ in Fig.~\ref{fig:bench}.
Due to the smallness of the parameter $\delta$, the positive exponential term is completely negligible close to the minimum where $\hat\phi=0$ and in most of the inflationary region. However for larger values of $\hat\phi$, this term will become important and spoil the flatness of the potential causing a departure from slow-roll.

As we shall see, the number of e-foldings depends on the range in field space where the inflaton potential can be kept flat, and so they will be inversely proportional to $\delta$. Larger numbers of e-foldings will then correspond to smaller values of $\delta$ which, in turn, imply a smaller value of $g_s$, and so a more fine-tuned situation. Therefore the underlying string construction is preferring a number of e-foldings close to the minimal value to solve the flatness and horizon problems, \textit{i.e.} $50\lesssim N_e\lesssim60$. This observation will be crucial later on in our study of the power loss at large angular scales.

Furthermore, the inflaton mass around the minimum and the uplifting contribution will set the Hubble constant during inflation when, due to the large value of $\hat\phi$, the two negative exponential terms are negligible:
\be
H_{\rm inf}\simeq \frac{V_{\rm inf}^{1/2}}{3} \simeq \frac{m_\phi}{2\sqrt{3}}\,.
\ee
If one then computes the effective inflaton mass during inflation:
\be
m^2_{\rm inf}\simeq -4 H_{\rm inf}^2 \,e^{-\frac{1}{\sqrt{3}}\,\hat\phi}\,,
\ee
it is easy to notice that it is tachyonic and that $|m_{\rm inf}^2|\ll H_{\rm inf}^2$ for large $\hat\phi$,
implying that this potential is naturally flat due to its typical exponential behaviour.

We finally mention that the expression \eqref{tau1norm} allows us to have a pictorial view of the
inflationary dynamics which takes place at constant volume with the inflaton slow-rolling towards its minimum
from large to small values of $\hat\phi$. This implies that during inflation the \emph{base} $\times$ \emph{fibre} structure
of the volume evolves from an initial situation with a large fibre $\tau_1$ and a small base $t_1\sim\tau_2/\sqrt{\tau_1}$
to a final situation with a small fibre and a large base keeping the overall volume constant.

\subsection{Slow-roll dynamics}
\label{SlowRoll}

We can now use the potential (\ref{InfPot}) to compute the slow-roll parameters $\epsilon$ and $\eta$ from \eqref{eqn:oldSR}. For $\delta\ll 1$ they turn out to be:
\bea
 \epsilon  &\simeq& \frac{8}{3} \left(
 \frac{  e^{-\frac{1}{\sqrt{3}}\hat\phi} -  e^{-\frac{4}{\sqrt{3}}\hat\phi}
 + \frac12 \, \delta \, e^{\frac{2}{\sqrt{3}}(1+n) \hat\phi}}
 {3 - 4  e^{-\frac{1}{\sqrt{3}}\hat\phi} +
  e^{-\frac{4}{\sqrt{3}}\hat\phi} + \frac{\delta}{(1+n)} \, e^{\frac{2}{\sqrt{3}}(1+n)\hat\phi}}
 \right)^2, \label{eps}\\
 \eta  &\simeq& -\frac{4}{3} \left( \frac{
 e^{-\frac{1}{\sqrt{3}}\hat\phi}
 - 4  e^{-\frac{4}{\sqrt{3}}\hat\phi} - \delta (1+n)\, e^{\frac{2}{\sqrt{3}}(1+n) \hat\phi}}
 {3 - 4  e^{-\frac{1}{\sqrt{3}}\hat\phi} +
 e^{-\frac{4}{\sqrt{3}}\hat\phi} + \frac{\delta}{(1+n)} \, e^{\frac{2}{\sqrt{3}}(1+n)\hat\phi}} \right)\,.
 \label{eta}
\eea
The inflationary dynamics can be understood by noticing that $\eta$ vanishes in two points, $\hat\phi_{\rm ip1}$ and $\hat\phi_{\rm ip2}$,
which determine three regions of field space, $\hat\phi\lesssim\hat\phi_{\rm ip1}$, $\hat\phi_{\rm ip2}\lesssim\hat\phi$ and $\hat\phi_{\rm ip1}\lesssim\hat\phi\lesssim\hat\phi_{\rm ip2}$:
\begin{itemize}
\item $\hat\phi\lesssim\hat\phi_{\rm ip1}$: for values of $\hat\phi$ close to the minimum, the positive exponential term can
be completely neglected, and so the slow-roll parameters take the simplified form:
\bea
 \epsilon  &\simeq& \frac{8}{3} \left(
 \frac{  e^{-\frac{1}{\sqrt{3}}\hat\phi} -  e^{-\frac{4}{\sqrt{3}}\hat\phi}}
 {3 - 4  e^{-\frac{1}{\sqrt{3}}\hat\phi} +
  e^{-\frac{4}{\sqrt{3}}\hat\phi}}
 \right)^2, \label{eps}\\
 \eta  &\simeq& -\frac{4}{3} \left( \frac{
 e^{-\frac{1}{\sqrt{3}}\hat\phi}
 - 4  e^{-\frac{4}{\sqrt{3}}\hat\phi}}
 {3 - 4  e^{-\frac{1}{\sqrt{3}}\hat\phi} +
 e^{-\frac{4}{\sqrt{3}}\hat\phi}} \right)\,.
 \label{eta}
\eea
Hence the potential develops an inflection point at $\hat\phi_{\rm ip1}$ where:
\be
\hat\phi_{\rm ip1}\simeq \frac{\ln 4}{\sqrt{3}}\simeq 0.80\,.
\ee
At this point $\epsilon$ becomes $\epsilon(\phi_{\rm ip1})\simeq 1.46$, and so the slow-roll
conditions are violated due to the vicinity to the minimum. However for larger values of $\hat\phi$,
$\epsilon$ decreases until slow-roll is established. This occurs rather close to the inflection point since
for $\hat\phi_{\rm end}=1$, $\epsilon$ becomes $\epsilon(\hat\phi_{\rm end})\simeq 0.78$ while $\eta(\hat\phi_{\rm end})\simeq -0.25$.
We shall consider this as a good point to end inflation (the final predictions are not sensitive to the exact point where inflation ends).

\item $\hat\phi_{\rm ip2}\lesssim\hat\phi$: for values of $\hat\phi\gg\hat\phi_{\rm ip1}$,
one of the negative exponential terms can be completely neglected, and the slow-roll parameters simplify to:
\bea
 \epsilon  &\simeq& \frac{8}{3} \left(
 \frac{  e^{-\frac{1}{\sqrt{3}}\hat\phi}
 + \frac12 \, \delta \, e^{\frac{2}{\sqrt{3}}(1+n) \hat\phi}}
 {3 - 4  e^{-\frac{1}{\sqrt{3}}\hat\phi} +
  \frac{\delta}{(1+n)} \, e^{\frac{2}{\sqrt{3}}(1+n)\hat\phi}}
 \right)^2, \label{epsS}\\
 \eta  &\simeq& -\frac{4}{3} \left( \frac{
 e^{-\frac{1}{\sqrt{3}}\hat\phi}
 - \delta (1+n)\, e^{\frac{2}{\sqrt{3}}(1+n) \hat\phi}}
 {3 +  e^{-\frac{4}{\sqrt{3}}\hat\phi} + \frac{\delta}{(1+n)} \, e^{\frac{2}{\sqrt{3}}(1+n)\hat\phi}} \right)\,.
 \label{etaS}
\eea
Thus the potential develops a second inflection point located at $\hat\phi_{\rm ip2}$ where:
\be
\hat\phi_{\rm ip2}\simeq -\frac{\ln \left[ \delta\left(1+n\right)\right]}{\sqrt{3}\left(1+\frac{2n}{3}\right)}
\approx -\frac{\ln \delta}{\sqrt{3}\left(1+\frac{2n}{3}\right)}\,.
\label{InflPoint2}
\ee
At this point $\epsilon$ is still satisfying the slow-roll condition since:
\be\label{eqn:ip2}
\epsilon(\phi_{\rm ip2})\simeq \frac 23\,\delta^{\frac{2}{3+2n}}\ll 1\,,
\ee
and the value of the spectral index is controlled by the small parameter $\delta$:
\be
n_s(\phi_{\rm ip2})-1\simeq 2\eta(\phi_{\rm ip2})-6\epsilon(\phi_{\rm ip2})\simeq -6\epsilon(\phi_{\rm ip2}) \simeq -4\,\delta^{\frac{2}{3+2n}}\,.
\label{nsblue}
\ee
As we shall see, $\delta$ controls also the number of e-foldings, creating a tension between obtaining enough e-foldings and, at the same time,
a correct value of the spectral index (in particular, a value of $n_s$ which is not too close to $1$ for $\delta\ll 1$).
If we now consider values of $\hat\phi>\hat\phi_{\rm ip2}$, parameterized as $\hat\phi=\hat\phi_{\rm ip2}+\hat\varphi$, $\eta$ changes sign
and becomes positive:
\be
\epsilon\simeq \frac{3}{8\left(1+n\right)^2} \,\eta^2  \qquad\text{where}\qquad
\eta  \simeq \frac49 (1+n) \delta^{\frac{1}{3+2n}} \, e^{\frac{2}{\sqrt{3}} (1+n)\hat\varphi}>0\,.
\ee
Clearly, $\eta$ still satisfies the slow-roll condition $\eta\ll 1$ if:
\be
\hat\varphi\ll \hat\varphi_{\rm max}\equiv-\frac{\ln\delta}{2\sqrt{3}(1+n)\left(1+\frac{2n}{3}\right)}\,.
\ee
However such values of the slow-roll parameters give a blue spectral index:
\be
n_s-1=2\eta-6\epsilon\simeq 2\eta >0\,.
\ee
As $\hat\phi$ increases even further, for $\hat\varphi\gg \hat\varphi_{\rm max}$,
the positive exponential dominates and the slow-roll conditions cease to be valid
since one obtains $\eta\simeq2\epsilon\simeq \frac 43(1+n)^2$. This takes place however still
within the regime of validity of the effective field theory \cite{Cicoli:2008gp}.

\item $\hat\phi_{\rm ip1}\lesssim\hat\phi\lesssim\hat\phi_{\rm ip2}$: in the region in field space between
$\hat\phi_{\rm ip1}$ and $\hat\phi_{\rm ip2}$ the scalar potential is dominated by
the $e^{-\frac{1}{\sqrt{3}}\hat\phi}$ term:
\be
 V_{\rm inf} \simeq \frac{m_\phi^2}{4}\left(3 - 4 \, e^{-\frac{1}{\sqrt{3}}\hat\phi} \right)\,, \label{SCALAR}
\ee
which gives rise to slow-roll parameters of the form:
\be
 \epsilon \simeq \frac{3 \,\eta^2}{2}\qquad \text{where}\qquad
 \eta \simeq -\frac 49 e^{-\frac{1}{\sqrt{3}}\hat\phi}\ll 1 \qquad\text{for}\qquad\hat\phi\gtrsim\hat\phi_{\rm ip1}\,.
\ee
The number of e-foldings $N_e$ can be worked out using the approximate potential (\ref{SCALAR}):
\be
 N_e=\int_{\hat\phi_{end}}^{\hat\phi_*}
 \frac{V_{\rm inf}}{V'_{\rm inf}} \; {\rm d} \hat\phi
 \simeq \frac{\sqrt{3}}{4} \int_{\hat\phi_{\rm end}}^{\hat\phi_*}
 \left( 3 \, e^{\frac{1}{\sqrt{3}}\hat\phi} -4 \right) \, {\rm d}\hat\phi
 = \left[ \frac94 \, e^{\frac{1}{\sqrt{3}}\hat\phi}
 - \sqrt3 \, \hat\phi \right]_{\hat\phi_{\rm end}}^{\hat\phi_*} \,,
 \label{Nefunc}
\ee
where $\hat\phi_*$ is the value of $\hat\phi$ at horizon exit whereas $\hat\phi_{\rm end}\simeq 1$
is the point where inflation ends. A good estimate for the maximum number of e-foldings
which is compatible with the observed value of the spectral index can be obtained by setting $\hat\phi_*=\hat\phi_{\rm ip2}$:
\be
N_e^{\rm max} \simeq \frac 94\, \delta^{-\frac{1}{3+2n}}+ \frac{3}{3+2n} \ln \delta+ \sqrt{3} -\frac 94\,e^{1/\sqrt{3}} \,,
\label{Nemax}
\ee
showing that more than 60 e-foldings requires $\delta \lesssim 3 \cdot 10^{-(5+3n)}$.
Expressing $\delta$ in terms of $g_s$ from (\ref{gs}), the previous expression for $N_e^{\rm max}$
can also be rewritten as (setting $C_1^{KK} =C_2^{KK}=C_{12}^W=\kappa_{122}=1$ which implies $\alpha = 1/\sqrt{2}$):
\be
N_e^{\rm max}\simeq \frac{k_1}{g_s^{k_2}}+3 k_2 \ln g_s+k_3\,,
\label{Negs}
\ee
where $k_1$, $k_2$ and $k_3$ are positive $n$-dependent parameters defined as:
\be
k_1\equiv \frac 94\ \frac{2^{-\frac{1+\frac{n}{3}}{3+2n}}}{(n+1)^{\frac{1}{3+2 n}}}\,,  \quad
k_2 \equiv \frac{4 (1+n/3)}{3+2 n}\,,\quad
k_3 \equiv \frac{3\ln\left[(1+n)2^{1+\frac{n}{3}}\right]}{3+2n}+ \sqrt{3} -\frac 94\,e^{\frac{1}{\sqrt{3}}}\,.  \nn
\ee
The expression (\ref{Negs}) clearly shows that the number of e-foldings is inversely proportional to the string
coupling to a positive power, and so smaller values of $g_s$ imply a larger $N_e$.
As an illustrative example, for $n=0$, $N_e^{\rm max}\gtrsim 50$ requires $g_s\lesssim 0.07$.
Therefore, situations with larger numbers of e-foldings are more fine-tuned than cases with smaller $N_e$.

This is also showing that if $\delta\simeq 3\cdot 10^{-5}$ for $n=0$ then $N_e=60$ for $\hat\phi_*=\hat\phi_{\rm ip2}$, but then from (\ref{nsblue})
we realize that $n_s(\phi_{\rm ip2})\simeq 1-4\,\delta^{2/3} \simeq -0.996$, giving a spectral index in tension
with cosmological observations. Thus $\delta$ should take smaller values so that horizon exit at $N_e=60$ occurs
well before the second inflection point.
\end{itemize}

\subsection{Predictions for the cosmological observables}
\label{ch:obsv}

The spectral index and the tensor-to-scalar ratio evaluated at horizon exit are given by:
\be
\label{rnsslowroll}
n_s = 1 + 2\eta_* - 6\epsilon_* \qquad \hbox{and} \qquad
r =16 \, \epsilon_* \,.
\ee
As we have seen, in the slow-roll region, we have $\epsilon_* = \frac32 \, \eta_*^2$
which implies the following typical prediction of this model:
\be
 r \simeq 6 (n_s - 1)^2 \,.
\label{corr}
\ee
Using suitable values of $\delta\lesssim 3 \cdot 10^{-(5+3n)}$ it is possible to obtain enough e-foldings
and correctly match the observed value for the spectral index $n_s\simeq 0.96\div 0.97$,
which in turn gives the prediction $r\simeq 0.005\div 0.01$.

Deviations from the correlation (\ref{corr}) arise for $\hat\phi_*$ close to $\hat\phi_{\rm ip2}$
where $\eta_* \simeq 0$ and $\epsilon_* \simeq \frac23 \, \delta^{\frac{2}{3+2n}}$, which lead to:
\be
 r \simeq \frac{32}{3} \, \delta^{\frac{2}{3+2n}}
 \qquad \hbox{and} \qquad
 n_s \simeq 1-4 \, \delta^{\frac{2}{3+2n}} \,.
\ee
In the extreme case $\delta \simeq 3 \cdot 10^{-5}$ for $n=0$ one obtains
$r \simeq 0.01$ and $n_s \simeq 0.996$.

The requirement of generating the right amount of density perturbations
fixes the overall normalization of the inflationary potential, and
so the value $m_\phi$ of the inflaton mass around the minimum.
The expression of the amplitude of the scalar perturbations is:
\be
 A_S\equiv\left(\frac{g_s}{8\pi}\right)
 \left(\frac{V_*^{3/2}}{V_*'}\right)^{2} \simeq 2.7\cdot 10^{-7}\,,
 \label{cobe}
\ee
where again the label $*$ denotes that the relevant quantities have to be evaluated at horizon exit,
and we have included the correct normalization prefactor $\left(g_s/8\pi\right)$
of the scalar potential obtained from dimensional reduction. This relation implies (setting $g_s\simeq 0.1$ and
restoring units of $M_P$):
\be
m_\phi^2\simeq 10^{-7} M_P^2\qquad\Leftrightarrow\qquad \langle\vo\rangle^{10/3}\simeq \frac{10^7}{g_s^{8/3}}\simeq 10^{10}
\qquad\Leftrightarrow\qquad \langle\vo\rangle\simeq 10^3\,.
\ee
In turn this gives an inflationary scale close to the GUT scale (restoring units of $M_P$):
\be
 M_{\rm inf}=V^{1/4}_{\rm end}\simeq \left(\frac{m_\phi}{M_P}\right)^{1/4} M_P \simeq
 10^{16}\,{\rm GeV} \simeq M_{\rm GUT}\,.
\ee

\section{Low Power at Large Scales}\label{ch:lowpower}

\subsection{Generalities}

Inflation predicts a nearly flat spectrum of primordial curvature perturbations. In terms of the angular power spectrum,
flat means that the Fourier coefficients should all be roughly normalized:
\be
\label{eqn:modenorm}
\frac{\ell(\ell+1)C_{\ell}}{6C_2}\approx 1\,.
\ee
The $C_{\ell}$ are defined as:
\be
\label{eqn:defCell}
C_{\ell} = \frac{1}{2\pi^2}\int_0^{\infty}dk\; k^2 \mathcal{P}(k)j_{\ell}^2(r_{\star}k)\,,
\ee
where $r_{\star}=14$ Gpc is chosen to set the scale of the power spectrum to the size of the universe at the surface of last scattering.
The flat spectrum of \eqref{eqn:modenorm} is achieved if $\mathcal{P}(k)$ scales precisely as $1/k^3$.
These primordial perturbations correlate directly to temperature fluctuations in the CMB.
Baryonic acoustic oscillations (BAO) populate this otherwise flat spectrum with harmonic peaks, the fundamental of which occurs near $\ell=200$. The Fourier modes, $C_{\ell}$'s, corresponding to the largest scales, $\ell\sim 2-20$, are not impacted by BAO effects, and so should be nearly constant.

Observations suggest that they are suppressed rather than constant. This was first observed by WMAP \cite{Bennett:2003bz}, and the significance of this anomaly has been strengthened by the Planck results \cite{Ade:2013zuv}. The $2\ell+1$ azimuthal modes associated to each $C_{\ell}$ can be considered as independent measurements, and therefore the theoretical uncertainty on the $C_{\ell}$ mode of the temperature variance measurements scales as $1/(2\ell+1)$.
This large theoretical uncertainty, known as ``cosmic variance'', may well explain the observations. If this result remains robust as more observations filter in, like a correlated distribution of tensor modes, alternative explanations may be more appropriate. Relating physics behind the power loss to other cosmological observables may allow for new model constraints and potentially a new window on primordial physics.
We explore this possibility in the context string inflationary models where the inflaton is a K\"ahler modulus,
focusing in particular on the case of Fibre Inflation.

Before turning to model details, let us examine how the inflaton dynamics can result in low power at large scales. One possible explanation for this anomaly arises from how inflation generates the primordial curvature perturbations. The Fourier modes of the perturbations do not evolve once they have exited the horizon. Once the acceleration has stopped, these modes reentered the horizon, starting with the most recently produced. In this sense,
looking at large scales is like looking backwards in time.

If the dynamics of the inflaton were such that a pivot from a blue to a red-shifted spectrum occurred just at the onset of the observationally relevant e-foldings of inflation, such an effect could explain the observed anomaly. Put differently, a large scale power loss may be the result of an early period of a blue-shifted spectrum. Of course, to agree with observations the average spectral index should be less than unity. Therefore, the pivot must have occurred sufficiently early.

A rolling scalar field can induce a pivot in the power spectrum by transitioning from one phase of dynamics to another.
The fast to slow-roll transition is a well studied example. This requires particular details of the initial conditions which are not well understood.
Fortunately, a pivot can also occur entirely within the framework of slow-roll inflation.
A slowly-rolling scalar whose trajectory undergoes a point of inflection will cause a pivot in the power spectrum. Since the scalar field tracks the logarithmic derivative of $V$ as in \eqref{eqn:newSR}, a point of inflection can be induced by a point of inflection in the potential as $\hat{\phi}_{\rm ip2}$ in \eqref{InflPoint2}.
This can occur either in large-field models like Fibre Inflation or in small-field models like inflection point inflation. In the former case, the pivot arises as a transition from a phase of power law to Starobinsky-like inflation. In the latter case, the transition is between chaotic and inflection point inflation.

There has been some suggestion that inflection point inflation cannot admit a sufficiently power pivot, owing to a ``symmetric'' property of the slow-roll parameter $\eta$. Nevertheless, inflection point inflation can support a strong, short period of a blueshifted inflation followed by a long, redshifted one. As discussed in \cite{Downes:2012gu}, starting from the slow-roll trajectory above the inflection point, there is a one-parameter family of solutions interpolating between these two extremes. The flaw in this argument lies in the use of the slow-roll parameter $\eta$. While the potential certainly satisfies these conditions, the dynamics \textit{does not}. Upon approach to the inflection point, the inflaton hops out of slow-roll temporarily, shifting from the dynamics of monomial-like potential to that of the inflection point. This point is important because precisely the same behavior occurs for the Fibre Inflation model discussed in this paper.

\subsection{Power loss from Fibre Inflation}

\subsubsection{The transition and the pivot}

The suppression of large scale fluctuations in the CMB can be explained by a pivot from a blue to a redshifted power spectrum near the onset of the observationally relevant perturbations. The existence of such a pivot can be explained naturally by the attractor dynamics of the gravity-scalar system. Essentially what is needed is a transition between two ``phases'' of inflation.

In the context of stringy models where the inflaton is a K\"ahler modulus, positive exponential terms in the potential give rise to a period of power law inflation at large inflaton VEVs. As the inflaton rolls towards the minimum of its potential, it transitions to a period of Starobinsky-like inflation. In this section we study this transition in the context of the Fibre Inflation model presented above and find that it illustrates some interesting general features.

\begin{figure}[h!]
\includegraphics{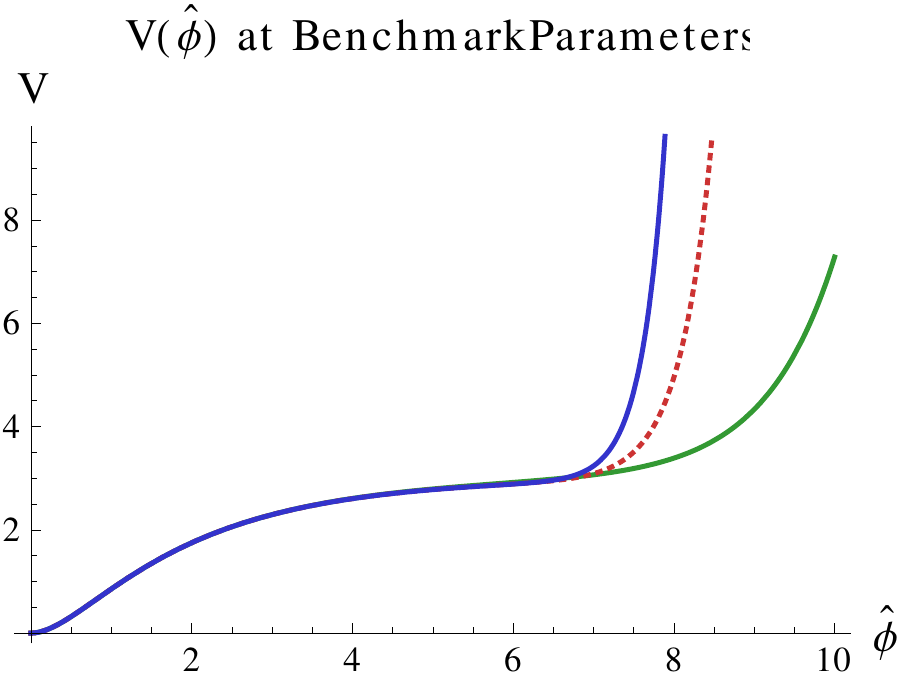}
\caption{The scalar potential $V_{\rm inf}$ as a function of $\hat{\phi}$ for three example points in the $(n,\delta)$-parameter space. In order from shallow to steep they are $(0,4.15\times 10^{-5}),(1.2,1.35\times 10^{-9})$ and $(2,3\times 10^{-12})$. The different values of $n$ were chosen to be below, near and above $n_c$, respectively. The value of $\delta$ were chosen to yield $N_e \sim 50$ redshifted e-foldings of inflationary expansion.}\label{fig:bench}
\end{figure}

To illustrate these features, we introduce the effect of the transition from a power law to a Starobinsky-like dynamics using three sets of parameters. Of interest will be the behavior below, near and above a ``critical'' value of $n$, $n_c$ defined below. The parameters will be $(n,\delta) = (0,4.15\times 10^{-5}),\,(1.2,1.35\times 10^{-9})$ and $(2,3\times 10^{-12})$. The potential $V_{\rm inf}$ for these parameter choices is illustrated in Fig.~\ref{fig:bench}.

Suppose the inflaton, $\hat{\phi}$, is rolling down the potential \eqref{InfPot}, from rest at some sufficiently high VEV. When $\hat{\phi}$ is above $\hat{\phi}_{\rm ip 2}$, the positive exponential dominates the dynamics and the field equation \eqref{eqn:coKGN} reduces to:
\be
\label{eqn:fastroller}
\hat{\phi}^{\prime\prime} \simeq \frac{1}{2}(\hat{\phi}^{\prime}-\sqrt{6})(\hat{\phi}^{\prime}+\sqrt{6})
\left[\hat{\phi}^{\prime}+\frac{2}{\sqrt{3}}\left(1+n\right)\right].
\ee
Solutions to \eqref{eqn:fastroller} asymptotically approach the middle of the three singular solutions. For power law inflation, this requires $\frac{2}{\sqrt{3}}(1+n)<\sqrt{6}$ for $n>0$, although accelerated expansion further requires that
$|\hat{\phi}^{\prime}|\approx\frac{2}{\sqrt{3}}(1+n)$ be less than $\sqrt{2}$.

\begin{figure}[h!]
\includegraphics{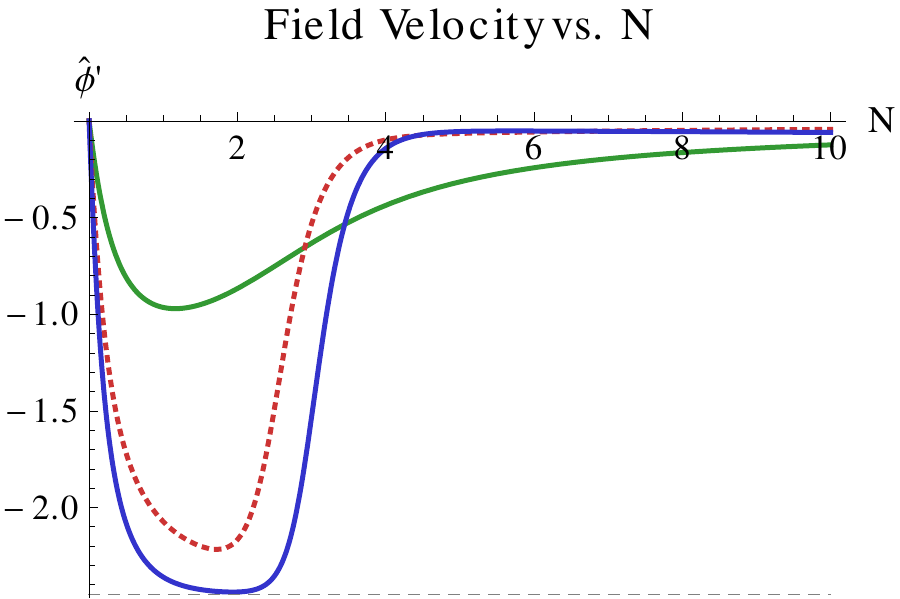}
\caption{$\hat{\phi}^{\prime}$ is shown as a function of $N_e$ for the same choice of parameters in Fig.~\ref{fig:bench}. A steeper potential leads to a sharper transition. The thin dashed line at $-\sqrt{6}$ represents a kinetic energy dominated solution. Note how the steepest potential induces a temporary period of fast-roll (decelerating) expansion before approaching a Starobinsky-like, slow-roll trajectory.}\label{fig:vels}
\end{figure}

Owing to the parametrization of time with $N_e$, $\phi^{\prime}=\pm \sqrt{6}$ corresponds to a period of kinetic energy domination. There is a critical value of $n$, $n_c=\frac{3}{\sqrt{2}}-1\simeq 1.12$, where the power law and the fast-roll solutions degenerate. For $n>n_c$, the inflaton will undergo a brief period of fast-roll, as seen in Fig.~\ref{fig:vels}.

\begin{figure}[h!]
\includegraphics{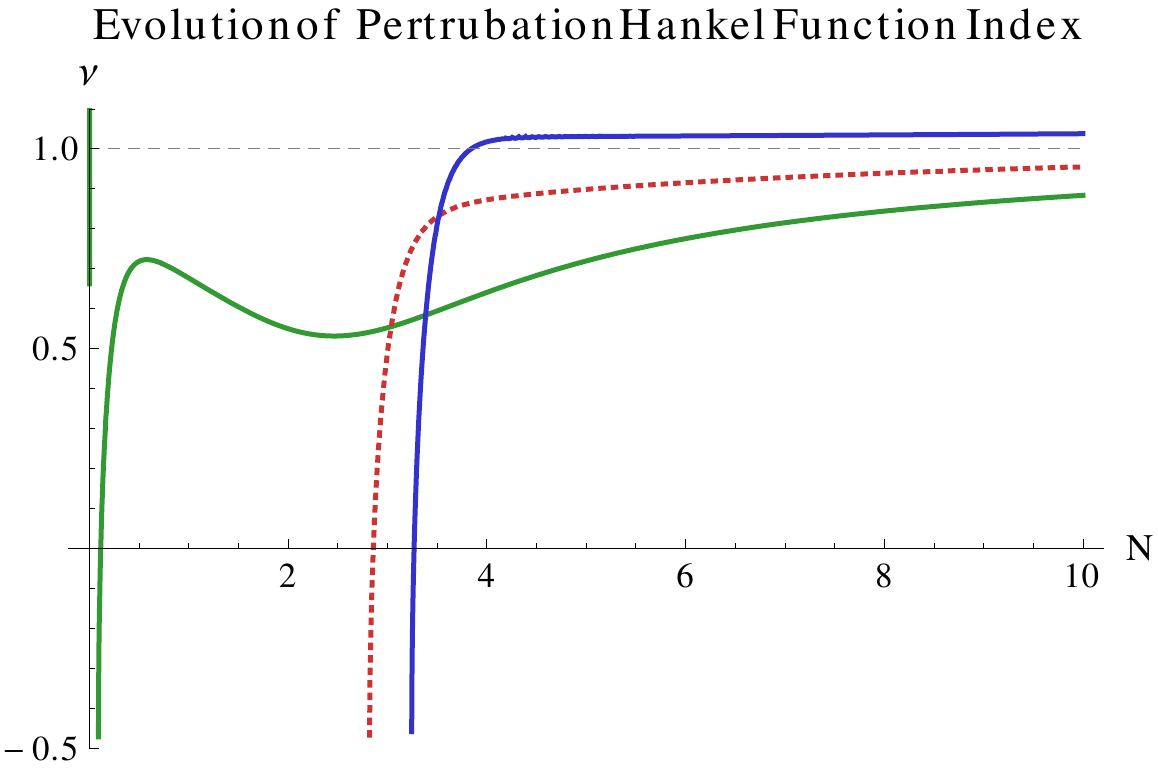}
\caption{The index of the spherical Hankel function $h^{(1)}_{\nu}$ associated to the inflaton wavefunction \eqref{eqn:BigQ} is plotted against $N_e$ for the same choice of parameters as Fig.~\ref{fig:bench}. $\nu=1$ corresponds to a scale-invariant spectrum. The two shallow cases ($n=0,1.2$) offer only a weak transition to a redshifted spectrum. The $n=2$ case is strong enough to induce a sharp transition, which becomes redshifted abruptly.}\label{fig:nus}
\end{figure}

This dynamics dominates while $\hat{\phi}\gtrsim \hat{\phi}_{\rm ip2}$, after which the slow-roll conditions from the remainder of the terms become relevant. For $n>n_c$, the brief period of fast-roll enhances the pivot from blue to redshifting, as seen in Figs.~\ref{fig:vels} and~\ref{fig:nus}.

\begin{figure}[h!]
\includegraphics{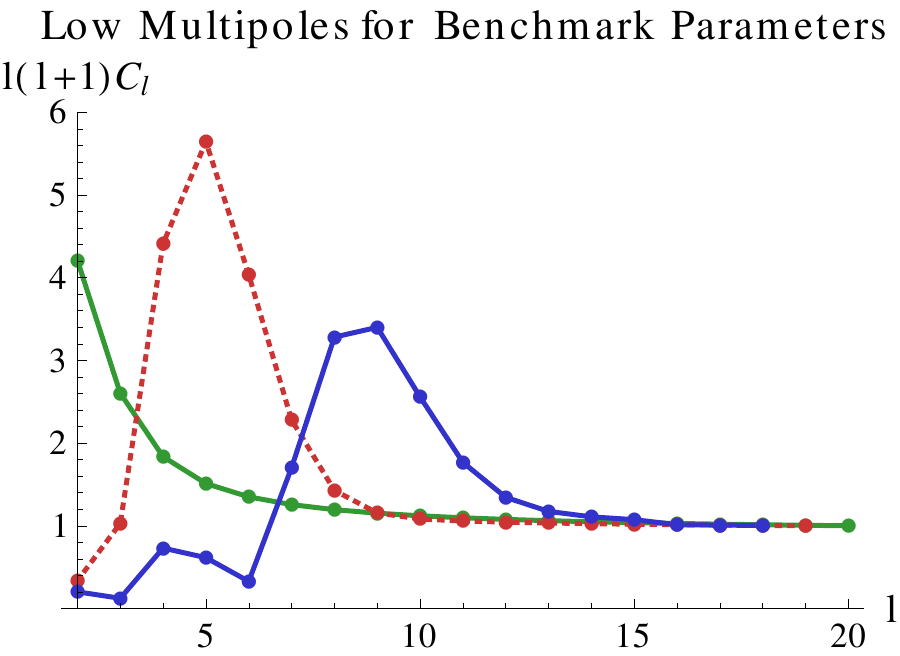}
\caption{The (normalized) angular power spectrum for a system starting from rest using the three parameter points from Fig.~\ref{fig:bench}. The scale $r_{\star}$ is chosen so that modes relevant for the CMB are affected just as the system pivots, leaving 50 e-foldings of redshifted inflation. The $n=0$ case behaves as standard models of inflation, with a redshifted spectrum. The $n=1.2,2$ cases show an initial suppression, followed by a peak, and finally an approach towards unity.}\label{fig:modes}
\end{figure}

If timed appropriately, this pivot can lead to the suppression of large scale fluctuations in the angular power spectrum, as illustrated in Fig.~\ref{fig:modes}. The suppression of power at large angular scales seen in Fig.~\ref{fig:modes} can be understood as follows. Each $C_{\ell}$ involves an integral over the comoving wavenumber $k$. Evaluated at horizon exit, the measure $k^2 dk$ becomes $(aH)^3(1-\frac{1}{2}\hat{\phi}^{\prime 2}) dN_e$. Therefore, at low $N_e$, i.e at low $k$, the amplitude of the power spectrum is suppressed by the velocity induced by the power law dynamics above $\hat{\phi}_{\rm ip2}$. This suppression is enhanced (and a peak is formed) by the norm of the wavefunction $|\mathcal{Q}_k|^2$. This is in part due to the evolution of the Hankel function index $\nu$, but also a spike in the value of $(aH\eta)$ near the transition.
Depending on the choice of parameters, such a well-timed suppression of power can interpolate between abrupt suppressing primarily the quadrupole, as with $n=1.2$, to somewhat extend, as with $n=2$.

\subsubsection{A phase space transition at $n_c$}

\begin{figure}[h!]
\includegraphics{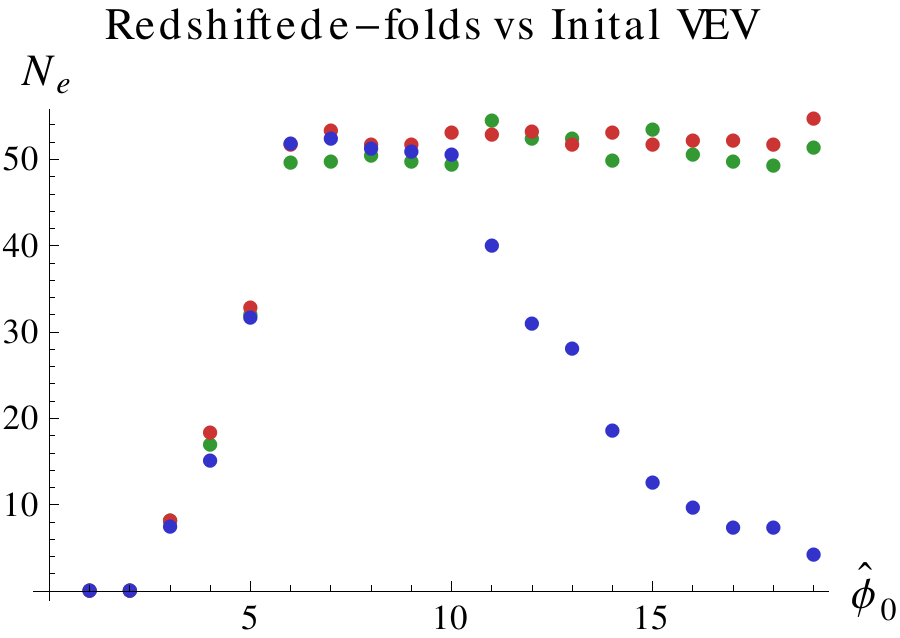}
\caption{A scan of initial values for $\hat{\phi}$ shows a different number of redshifted e-foldings, inclusive of the three parameter choices from Fig.~\ref{fig:bench}. The behavior is universal for low initial VEV, as the slow-roll trajectories are essentially the same --- Starobinsky-like --- below $\hat{\phi}_{\rm ip2}$. For both of the shallow trajectories, the number of redshifted e-foldings levels off for large initial VEVs. In the steep ($n=2$) case, the value of $N_e$ decreases for increasing VEVs. This fall off becomes more pronounced as $n$ grows above $n_c$, and arises from an ``overshoot'' of $\hat{\phi}_{\rm ip2}$; $\hat{\phi}$ transitions to the slow-roll trajectory later and later for increasing $n$.}\label{fig:dots}
\end{figure}

Let us consider a fixed value of $\delta$. For $n<n_c$, all initial VEVs of $\hat{\phi}$ lead to the same number of redshifted e-foldings of inflation. Increasing the initial VEV just prolongs the duration of the power law expansion. Above $n_c$, however, a larger initial VEV leads to fewer redshifted e-foldings. The reduction in inflationary e-foldings occurs because the transition from fast-roll evolution to slow-roll inflation occurs ever further past $\hat{\phi}_{\rm ip2}$. In other words, $\hat{\phi}$ overshoots the inflection point. This is illustrated in Fig.~\ref{fig:dots}.

We stress again that this is all due to the attractor dynamics of the gravity-scalar system. Indeed, this ``overshooting'' behavior of the transition is reminiscent of the fixed point behavior of inflection point inflation \cite{Itzhaki:2008hs}. Here the authors studied a one-parameter family of inflection point models and found a critical value of the coupling for which the inflaton VEV asymptotically approaches the inflection point. By varying the coupling near its critical value, it was found to have the properties reminiscent of a second order phase transition. In \cite{Downes:2012xb}, it was shown that this phenomenon generalizes to all of phase space and general inflection point models. Given the similarity to the Fibre Inflation case, it is tempting to conjecture that these two models  belong to the same universality class. At least to the extent that the analogy with statistical mechanics holds.

There are notable differences between the two cases, however. The inflection point case occurs with infinite, blueshifted e-foldings. This stands in contrast to the finite, redshifted case of Fibre Inflation. These distinctions are ameliorated in the physical case of a deformed inflection point with finite e-foldings, which suggests a relationship between the inflection point deformation parameter and the $\delta$ parameter of Fibre Inflation.
It is interesting to observe that the $\delta=0$ case of Fibre inflation corresponds in some sense to an inflection point model with an infinitely (rather than doubly) degenerate critical point --- the potential is asymptotically flat. We leave further analysis of such considerations for future work.

\subsubsection{Phenomenologically viable parameters}

The potential $V_{\rm inf}$ from \eqref{InfPot} has two parameters $\delta$ and $n$. Taken with the initial position and velocity of $\hat{\phi}$, there are four parameters in total in the model.
The exact value of $n$ depends on the underlying string loop dynamics, and can be thought of as a scaling dimension for the inflaton.
The parameter $\delta \sim g_s^{4+\frac{4n}{3}}\ll 1$ is the small number typically required by inflationary models to achieve sufficiently many e-foldings of expansion. An interesting feature of this model is that $\delta$ becomes automatically very small when one requires $g_s\ll1$ in order to trust the effective field theory. Taken together these parameters give rise to a delicate balance of exponential terms in the effective potential,
which is a fairly generic situation in string inflationary models based on K\"ahler moduli \cite{Cicoli:2011zz}.

Because the mechanism presented above to suppress large scale perturbations is tied to the attractor dynamics of the inflaton,
the details of the initial inflaton velocity contribute nothing beyond what was studied in \cite{Downes:2012xb}. Indeed, the original kinetic-energy dominated scenario studied in \cite{Contaldi:2003zv} is generated automatically. Therefore, concerns about tuning the initial conditions --- of which we have little conceptual understanding --- can be ported to concerns about tunings in the potential.

As discussed above, there are some issues with the initial value of the inflaton. At least to the extent considered here, any restrictions on the initial $\hat{\phi}$ are degenerate with the choice of $\delta$. Given a full model of inflation and reheating, it may be possible to break this degeneracy, perhaps with the precise value of $n_s$ and its running.

\begin{figure}[h!]
\includegraphics[scale=0.8]{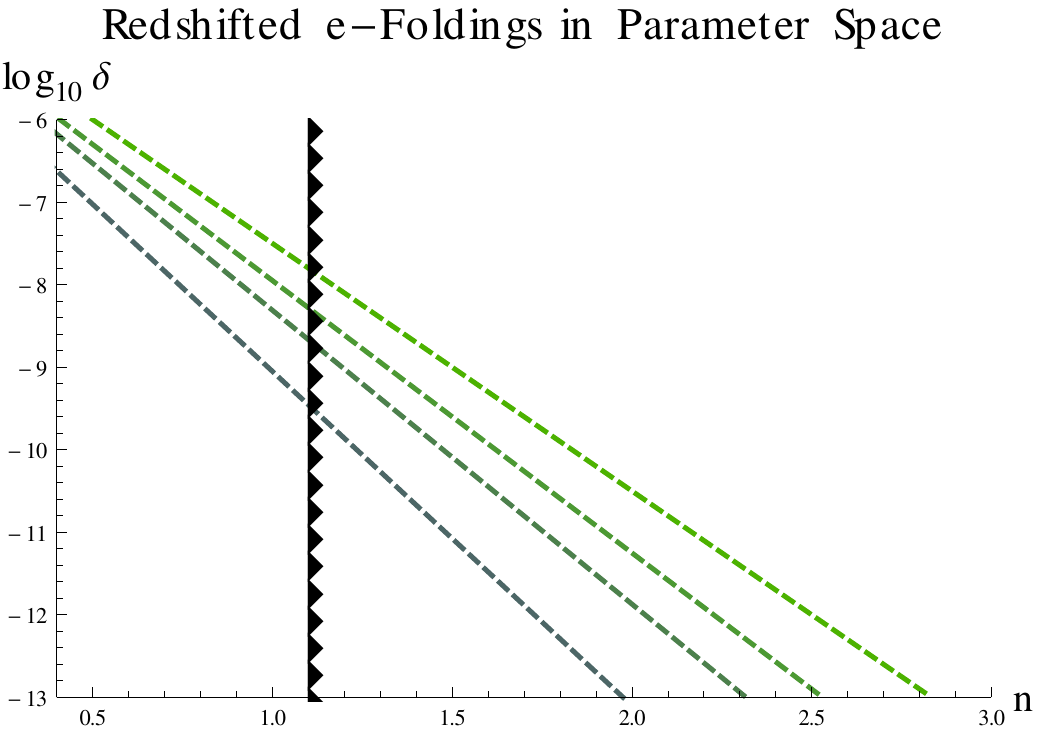}
\caption{The number of redshifted e-foldings of inflation is shown in the $(n,\log_{10} \delta)$ parameter space. The line of arrows represents $n_c$, and points to the region with fast-roll behavior. The dashed lines are contours of constant e-foldings, at $40, 50, 60$ and $100$ in descending order from the top right. Sufficient power loss and the requisite number of e-foldings limit the Fibre Inflaton model to the strip on the right side bounded by these contours. Note that the precise value of $N_e$ depends on a number of model details like, for example, the systematics of reheating.}\label{fig:paras}
\end{figure}

In Fig.~\ref{fig:paras} the number of redshifted e-foldings of inflation is illustrated in the $(n,\delta)$-plane. The phenomenologically favored region of $40 \lesssim N_e \lesssim 60$ with $n>n_c$ is shown as a bounded region. The three example points used in the Figures above are also highlighted.

We end this section with a final comment on initial VEVs. The fast-roll period of inflation cannot happen over a large region of field space since $V$ is bounded above by the Kaluza-Klein scale. As a result, a larger initial $\hat{\phi}$ strongly constrains the overall scale of the potential. This is useful given that the potential of a canonical scalar field enters the equations of motion via its logarithmic derivative \eqref{eqn:coKGN}. Given the finite range of $\hat{\phi}$ allowed when $n>n_c$ (see Fig.~\ref{fig:dots}), this may lead to a correlation between power loss and the scale of inflation. We leave this interesting matter for future work.

\section{Discussion}\label{sec:five}

The observation of a slightly redshifted spectrum of temperature fluctuations gives a strong constraint on models which attempt to describe the early universe. Moreover, the small but persistent evidence for a suppression of power at large angular scales warrants a theoretical mechanism.
This mechanism can be based on the primordial dynamics of the background which generates an appropriately timed pivot from a blue to a redshifted power spectrum.

While a definitive observation requires a considerably enhancement in statistical significance --- which is non-trivial to attain
given considerations of cosmic variance --- the imprint of a primordial semi-classical dynamics of the inflaton on the CMB is an intriguing possibility.

In this work we have explored this possibility using scalar potentials derived from string models of inflation based on K\"ahler moduli \cite{Cicoli:2011zz}.
What we have found mirrors the inflection point case studied in \cite{Downes:2012gu}: a pivot in the power spectrum is generated when the inflationary dynamics undergoes a transition.
In the present case, this transition is to a Starobinsky-like slow-roll phase from a fast-roll period generated by power law dynamics.
The pivot is strong enough only if the field velocity during the power law phase
--- dictated by the coefficient of $\hat{\phi}$ in the dominant exponential term --- is very large.
In other words, we have given a concrete realization of the scenario proposed by \cite{Contaldi:2003zv}.
Just as the standard arguments for inflation, this scenario enjoys the stability afforded by the attractor dynamics.

By focusing on the case of Fibre Inflation \cite{Cicoli:2008gp} and 
by parameterizing the potential in terms of a scaling dimension $n$ and a tuning parameter $\delta$,
we investigated how this transition can give rise to large scale power suppression.
We found a critical value $n_c$ above which a pivot in the power spectrum can be sufficiently sharp.
Above $n_c$, the space of possible initial conditions also changes,
as the fast-roll phase of the inflaton tends to overshoot the Starobinsky-like region of the potential.
This gives an interesting parallel to the work of \cite{Itzhaki:2008hs}. The r\^ole of $\delta$ in this analysis is the standard ``small number'' required to achieve the appropriate number of inflationary e-foldings.

Importantly, $\delta$ scales with a positive and large power of the string coupling, as seen from \eqref{gs}. The validity of the effective field theory requires $g_s$ to be small, and therefore $\delta$ is also suppressed. While this affords a sizable number of e-foldings,
$g_s$ does not need to be extremely fine-tuned. For the case of $n=2$, fifty redshifted e-foldings amounts to $\delta\sim g_s^{20/3}\sim 10^{-11}$,
suggesting $g_s\sim 10^{-33/20}\sim 0.02$. Such a number is fairly small but not unreasonable. Since larger $N_e$ would require an even greater tuning in $\delta$, any ``coincidence problem'' with respect to the timing of the pivot can be reinterpreted as a naturalness condition on $g_s$.

A by-product of this analysis is an exception to the conventional wisdom surrounding various models for power suppression and redshifting.
Owing to the cubic nature of the potential near the inflection point, it is often argued that (\textit{i}) twice the observed number
of e-foldings are required to ensure they are redshifted and because of this (\textit{ii}) significant power loss from inflection point models is impossible to achieve. Moreover, ($\textit{i}$) indicates that observations require a much strong flattening --- and therefore tuning --- of the potential to achieve sufficient e-foldings. These claims are based on the fact that the slow-roll parameter $\eta$ is ``symmetric'' about the inflection point.

This suggests that alternative scenarios, like Fibre Inflation, would give an enhanced ability to generate such power loss due to the fact that $\eta$ is ``asymmetric'' in the inflationary region. $\eta$ is quite asymmetric for $n=0$, which corresponds to the standard case of Fibre Inflation and the power spectra look surprisingly close to the inflection point case. Low power requires something else entirely.

The symmetric-asymmetric $\eta$ argument fails because the slow-roll approximation breaks down during a transition from one phase of inflation to another. For the pivot --- and therefore loss of power --- to be sharp, the inflaton velocity must change abruptly. In the inflection point case, this is obtained near an attractor phase transition (and more generally near the boundary of the basin of attraction). For the Starobinsky-like case considered in this work, it is induced by the abrupt exit from a period of fast-roll power law dynamics. Note also that a sharp transition, as seen with the $n=2$ case above, markedly reduces the required total number of e-foldings to achieve the expected $N_e \sim 50$ or so of redshifted expansion.

It is interesting that the background evolution of the parameter $\Xi$ and the wavefunction $\mathcal{Q}_k$ are remarkably similar for both the Starobinsky-like and inflection point cases. We intend to further investigate this relationship in future work.

\textit{Note added: This paper is submitted simultaneously to the related work \cite{PW}.}

\begin{acknowledgements}
This work is supported in part by the DOE grant DE-FG02-95ER40917. We are grateful to Cliff Burgess, Fabio Finelli, Alessandro Gruppuso,
Marco Peloso and Fernando Quevedo for interesting discussions and constructive comments.
\end{acknowledgements}

\newpage
\bibliography{lowp}

\end{document}